# Fast switchable unidirectional magnon emitter

*Yueqi Wang[1,#], Mengying Guo[1,#], Kristýna Davídková[2], Roman Verba[3], Xueyu Guo[1], Carsten Dubs[4], Andrii V. Chumak[2], Philipp Pirro[5], Qi Wang[1*]*

[1] *School of Physics, Hubei Key Laboratory of Gravitation and Quantum Physics, Institute for Quantum Science and Engineering, Huazhong University of Science and Technology, Wuhan, China*

[2] *Faculty of Physics, University of Vienna, Vienna, Austria*

[3] *Institute of Magnetism, Kyiv, Ukraine*

[4] *INNOVENT e.V., Technologieentwicklung, Jena, Germany*

[5] *Fachbereich Physik and Landesforschungszentrum OPTIMAS, Rheinland-Pfälzische Technische Universität Kaiserlautern-Landau, Kaiserslautern, Germany*

**Abstract:**

Magnon spintronics is an emerging field that explores the use of magnons, the quanta of spin waves in magnetic materials for information processing and communication. Achieving unidirectional information transport with fast switching capability is critical for the development of fast integrated magnonic circuits, which offer significant advantages in high-speed, low-power information processing. However, previous unidirectional information transport has primarily focused on Damon-Eshbach spin wave modes, which are non-switchable as their propagation direction is defined by the direction of the external field and cannot be changed in a short time. Here, we experimentally demonstrate a fast switchable unidirectional magnon emitter in the forward volume spin wave mode by a current-induced asymmetric Oersted field. Our findings reveal significant nonreciprocity and nanosecond switchability, underscoring the potential of the method to advance high-speed spin-wave processing networks.

[#] These authors contributed equally to this work
[*] Corresponding Author: williamqiwang@hust.edu.cn

**Introduction**

Spin waves, or their quanta magnons, are collective excitations in magnetic materials that can be utilized for the transfer and processing of information [1-4]. Spin waves have recently attracted considerable attention due to their distinctive properties, including Joule-heat-free transfer [1], rich nonlinear effects [6-9], down to nanoscale wavelengths [10-15] and up to THz operation frequency [16-18]. In light of these properties, there has been a great deal of extensive research conducted on magnonic devices, including magnonic logic gates [19-21], transistors [20-22], directional coupler [23-25] and Y-shape combiners [26].

Unidirectional information transport is an intriguing and often crucial phenomenon for information processing systems, which attracts attention in optical [27,28] and microwave bands [29], as well as in electronics [30-32]. Magnonics offers unique inherent abilities for unidirectional information transport, most of which relies on the spin wave nonreciprocity - phenomenon when dispersion curve, dissipation or other propagation characteristics of oppositely propagating waves are not the same [33]. By a proper design it is possible to ensure unidirectional spin wave propagation at certain frequencies or frequency band.

Previous studies have focused on spin waves in the so-called Damon-Eshbach (DE) geometry [34], when waves propagate perpendicularly to the static magnetization direction which, in, turns, is in-plane. In DE geometry frequency and/or dissipation nonreciprocity is realized in many different systems: ferromagnetic bilayers and synthetic antiferromagnets [35,36], ferromagnetic-heavy metal bilayer exhibiting interfacial Dzyaloshinskii–Moriya interactions [37,38], simply metalized film made of ferromagnetic dielectric [34] or saturation magnetization gradients [39]. Also, chirality of DE spin waves offers excitation nonreciprocity in a simple ferromagnetic film by a microstrip antenna or nanoscale grating [40,41]. However, in all the mentioned systems the direction of the information transport is not switchable, i.e., the propagation direction is determined once the geometry and external field are fixed and cannot be switched in a short time.

Recently, the forward volume spin waves (FVSWs) in narrow waveguide have shown several novel features, such as large stable precession angle, deeply nonlinear shift, huge bistable window and possible to switch between them [42-45], offering advantages for integrated magnonic circuits. At the same time, FVSWs are isotropic, i.e., the dispersion curve is symmetric and their propagation characteristics are the same in all directions, so the previous method to realize the unidirectional information transport is not suitable for FVSWs modes.

Here, we present a method to achieve unidirectional spin-wave excitation by controlling the energy landscape under and near the excitation antenna, without introducing nonreciprocity into the spin-wave dispersion. Switching of the propagation direction is achieved by a modification of the energy landscape and does not rely on the magnetization reversal (as in DE case), thus, can be done within nanoseconds. This fast-switching unidirectional magnon emitter is realized by generating an asymmetric Oersted field by passing a direct current (DC) through the same source antenna. Oersted fields create a barrier for spin waves in one direction within a certain bandgap, proportional to DC. Unidirectional magnon emission occurs for spin-wave frequencies within this gap, and the excitation direction can be reversed by DC polarity. The nanosecond switching capability of this emitter is critical for the development of high-speed magnonic networks.

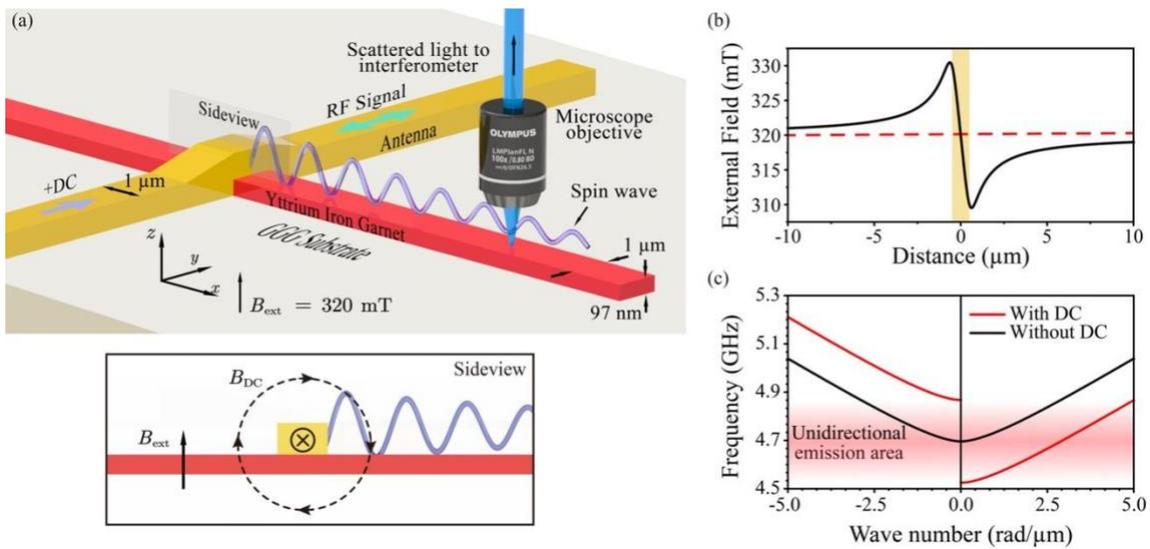

*Figure 1. **Schematic representation of the unidirectional magnon emitter.** (a) Sketch of the sample and experimental configuration: A 1 μm wide antenna is placed on a 1 μm wide and 97 nm thick yttrium-iron-garnet (YIG) waveguide grown on a gadolinium-gallium-garnet (GGG) substrate. The intensity of the spin waves is measured using microfocused Brillouin light scattering spectroscopy (μBLS). The unidirectional excitation is a consequence of the Oersted field generated by the DC signal, as illustrated in the sideview. Changing the direction of the DC signal reverses the nonreciprocal propagation direction. (b) The total external magnetic field (black line) is the combined effect of a 320 mT static magnetic field (red dashed line) and the Oersted field generated by the 30 mA DC at the antenna (yellow rectangle). (c) The dispersion curve of FVSWs under uniform external field of 320 mT (black). Red curves show dispersion curves, shifted by Oersted fields, for positive (negative) k dispersion is calculated using maximum and minimum Oersted fields on the left (right) side of antenna. A pink area indicates the unidirectional excitation emission region.*

**Results**

**General concept of the unidirectional magnon emitter.** Figure 1(a) is a schematic diagram of the experimental setup for the unidirectional magnon emitter. A 1 μm wide Au antenna is placed on a 1 μm wide and 97 nm thick yttrium-iron-garnet (YIG) waveguide for spin-wave excitation. An out-of-plane external field of 320 mT is applied by a NdFeB permanent magnet to saturate the magnetization along the z-axis. Radio frequency (RF) of -5 dBm power is applied to the antenna to excite the propagating spin waves. Microfocused Brillouin light scattering spectroscopy (μBLS) is used to detect the spin-wave intensity at a location around 3 μm away from the antenna on both sides. In general, the FVSWs are isotropic and the dispersion curves on both sides of the antenna are the same (Fig. 1(c) black line). However, when a DC signal is applied to the antenna, the current-induced Oersted field ($B_z$) on both sides of the antenna is asymmetric, as it is shown schematically in the sideview of Fig. 1(a) (bottom). The black curve in Fig. 1(b) illustrates the total external magnetic field combined with a uniform external field (320 mT, red dashed line) and a DC-induced Oersted field, which is asymmetric on both sides of the antenna. The asymmetric field creates a potential barrier on one side of the antenna and a potential well on the other side resulting in different shift (position-dependent) of spin-wave dispersion curves on both sides. Figure 1(c) shows maximal shifts of the dispersion curve at the maximum and minimum of Oersted fields (schematically, we show up (down) shifted dispersion at negative (positive) $k$). Spin wave propagation in the negative direction (to the left of antenna) is constrained or even completely blocked when spin wave frequency is below the minimum of the shifted dispersion, while no obstacles take place for the propagation in the opposite direction (the minimal allowed for propagation frequency is even smaller than those of an unperturbed structure). Thus, there is frequency band with nonreciprocal excitation, which could become completely unidirectional if the barrier is sufficiently high and/or large to prevent tunnelling effects.

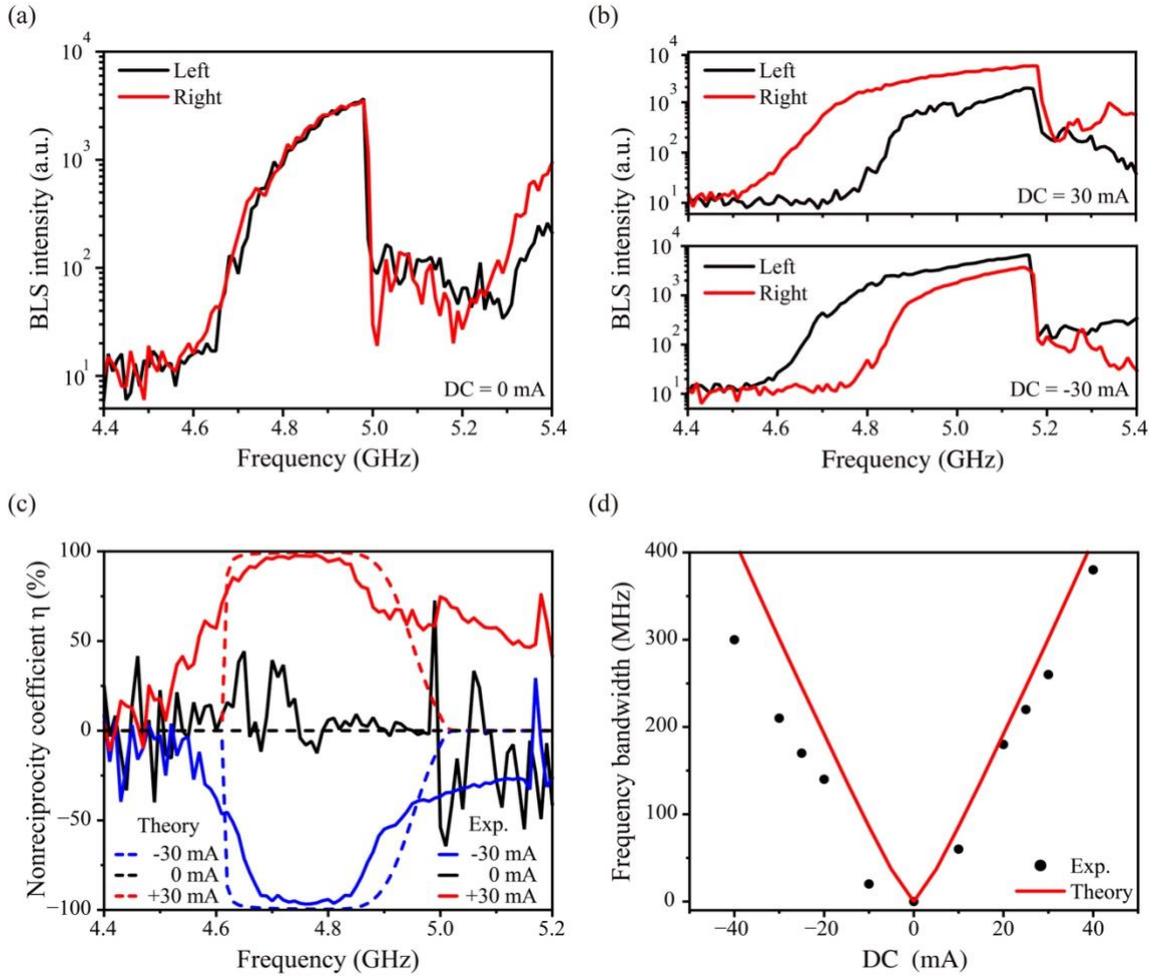

*Figure 2. **Nonreciprocity of the spin wave in frequency sweep curves**. Microwave signals of different frequencies are applied with the addition of a DC signal. The intensity of the spin waves at the symmetric position on the waveguide relative to the antenna is measured using μBLS. (a) In the absence of DC, the transmission curves are consistent on both sides, indicating no nonreciprocity. (b) Top: the propagation of spin waves in the waveguide under the influence of 30 mA DC is nonreciprocal in a certain frequency range. A frequency gap is observed between the frequency sweep curves on both sides. Bottom: the reverse current of -30 mA reverses the nonreciprocity, giving an experimental result opposite to that of 30 mA. (c) The experimental (solid lines) and analytical (dashed lines) nonreciprocity coefficient as a function of frequency for different conditions. (d) The frequency bandwidth of the nonreciprocity, where the nonreciprocity coefficient $|\eta| > 75\%$ for different DC magnitudes (black dots) and theoretical calculation (red lines).*

To excite spin waves, pulsed microwave signals are applied to the antenna, sweeping from 4.4 GHz to 5.4 GHz with a frequency interval of 10 MHz. The microwave power is fixed at -5 dBm with a pulse duration of 3 μs and a period of 4 μs. The focused laser spot of μBLS is positioned at symmetric locations around 3 μm on each side of the waveguide relative to the antenna to measure the spin-wave intensity as a function of frequency. Figure 2(a) shows a typical FVSWs excitation spectrum, i.e., the spin-wave intensity increases with increasing frequency and suddenly drops to a thermal level around 5.0 GHz, showing a foldover effect [42-46]. In addition, Fig. 2 shows a small signal increase above 5.2 GHz due to the direct excitation of higher spin-wave width modes in the lateral confinement waveguides [43,47,48]. As expected, Fig. 2(a) shows similar spectra on both sides of the antenna with no DC applied indicating an isotropic (reciprocal) excitation of the FVSWs.

Figure 2(b) shows the main results of this research - the spin-wave excitation spectra becomes nonreciprocal when a DC of 30 mA is applied (top panel). On the left side of the antenna, the spectrum is similar to the previous one, i.e., the spin-wave signal starts to increase from 4.6 GHz. But on the right side, the starting point shifts to a higher frequency of about 200 MHz. Thus, a unidirectional excitation frequency gap appears with a frequency range from 4.6 GHz to 4.8 GHz. As discussed earlier, the unidirectional emission is caused by the asymmetric Oersted field induced by the DC passing through the antenna. The asymmetric field creates a potential barrier on the left side of the antenna and a potential well on the right side as shown in Fig. 1(b). The potential barrier can completely block the low-frequency spin wave propagation and also cause partial reflection for the high-frequency spin waves, so that the spin-wave intensity of the black curve is lower than that of the red curve even outside the unidirectional frequency gap. On the other side, the spin waves pass through the potential well by wavelength conversion [43], thus realizing the unidirectional magnon emitter. Please note that the jump frequency, where the spin-wave intensity suddenly drops, also increases from about 5.0 GHz to 5.2 GHz when the DC signal is applied. A similar phenomenon is observed in micromagnetic simulation without the inclusion of Joule heating [see Supplementary Materials]. The intensity drop in a nonlinear excitation is known to be determined by the balance of drive power and total losses [43]. Thus, we conclude that the higher frequency position of the drop is a consequence of the reduction of the total radiative losses (in both directions) under applied Oersted fields. Radiative losses are defined by the spin-wave group velocities (in our case, by a functional of the position-dependent group velocities), and the group velocity decrease in the negative direction is higher by modulus than the increase in the positive direction because of almost quadratic spin-wave dispersion relation.

To experimentally evaluate the effect of Joule heating, a reverse DC is applied. The bottom panel of Fig. 2(b) shows the measured spin-wave excitation spectra at the applied DC of -30 mA. This nonreciprocity is reversed when the current direction is reversed. Therefore, the observed nonreciprocity is mainly caused by the asymmetric Oersted field rather than Joule heating, since the nonreciprocity induced by thermal effects should be independent of the current direction. This is also confirmed by micromagnetic simulations. In the supplementary materials, the similar results of Fig. 2(a-c) are repeated by micromagnetic simulations without considering the Joule heating. This

Oersted field-induced nonreciprocity paves the way for the realization of the fast-switching property and will be discussed later.

To describe the nonreciprocity quantitatively, we define the nonreciprocity coefficient as follows $\eta = \frac{I_r - I_l}{I_r + I_l}$, where $I_r$ and $I_l$ represent the spin-wave intensities on the right and left sides of the antenna, respectively. Figure 2(c) shows the relationship between the nonreciprocity coefficient $\eta$ as a function of the spin-wave frequency with DC of 0 mA and $\pm 30$ mA. The interesting frequency range is from 4.6 GHz to 5 GHz, where the spin waves are fully excited in all cases. The coefficient $\eta = 0\%$ represents a uniform excitation for both sides of the antenna. Conversely, $\eta = \pm 100\%$ indicates perfect unidirectional magnon excitation for the left/right propagating spin waves. The maximum nonreciprocity coefficient is close to ±100% at the frequency around 4.75 GHz and a wide frequency bandwidth of 200 MHz from 4.65 GHz to 4.85 GHz with coefficient $|\eta| > 75\%$ is observed for both DC of ±30 mA. As expected, no nonreciprocity is observed in the absence of a DC signal. Additionally, we tested the frequency sweep curves under currents ranging from 0 to ±40 mA and calculated the nonreciprocity [see Supplementary Materials].

In order to qualitatively describe experimental findings, we develop a model focusing on the impact of DC Oersted fields on spin-wave excitation nonreciprocity and, thus, ignoring all the nonlinear effects. The dispersion relation in the presence of spatially varying Oersted fields is given by the linear relation

$$\omega_k(\Delta B) = \omega_{k,0} + \gamma \Delta B, \tag{1}$$

where $\omega_{k,0}$ is the dispersion in the absence of Oersted field, and $\Delta B(x)$ represents the position-dependent Oersted field. When spin wave frequency $\omega$ is below the instant FMR frequency, it attenuates with the rate

$$\kappa(x) \approx \sqrt{\frac{\omega_0 + \gamma \Delta B(x) - \omega}{\omega_M \lambda^2}}, \tag{2}$$

where $\lambda$ is the exchange length, while for a wave having frequency above instant dispersion minimum the attenuation is defined by the damping $\Gamma$ and instant group velocity $v_{gr}$, $\kappa = \Gamma/v_{gr}$. The total power attenuation rate for counter-propagating spin wave is given by

$$A_\pm = \int_0^L e^{-2\kappa[\Delta B(\pm x)]x} \, dx, \tag{3}$$

where $L$ is the distance to the measurement point (3 μm in our case). To get nonreciprocity coefficient, we also account for the frequency-dependent excited power $P_0(\omega)$ and thermal level $P_{kT}$, so that spin-wave power at the measurement points is $P_\pm = P_0 A_\pm + P_{kT}$. The details of the derivation can be found in Supplementary Materials.

The dashed lines in Fig. 2(c) show the analytical curves of the nonreciprocity coefficient as a function of frequency, which correspond well with the experimental results in the frequency range of interest. The deviation is observed in the low and high frequency range due to several effects, ignored in the model, namely, the higher width mode excitation, nonlinear effects, and weak wave reflection when its energy exceeds the barrier. Figure 2(d) summarizes both the experimental (black dots) and theoretical (red line) frequency bandwidths of the nonreciprocity (where $|\eta| > 75\%$ in the interesting frequency range from 4.6 GHz to 5.0 GHz) for different applied currents. Higher DC currents lead to a higher potential barrier, ultimately resulting in a wider frequency bandwidth. The model, being quite a simplified as described above, overestimates the nonreciprocal bandwidth, but not much, indicating that the spin wave tunneling and decay under the barrier are primary responsible for the nonreciprocity of excitation.

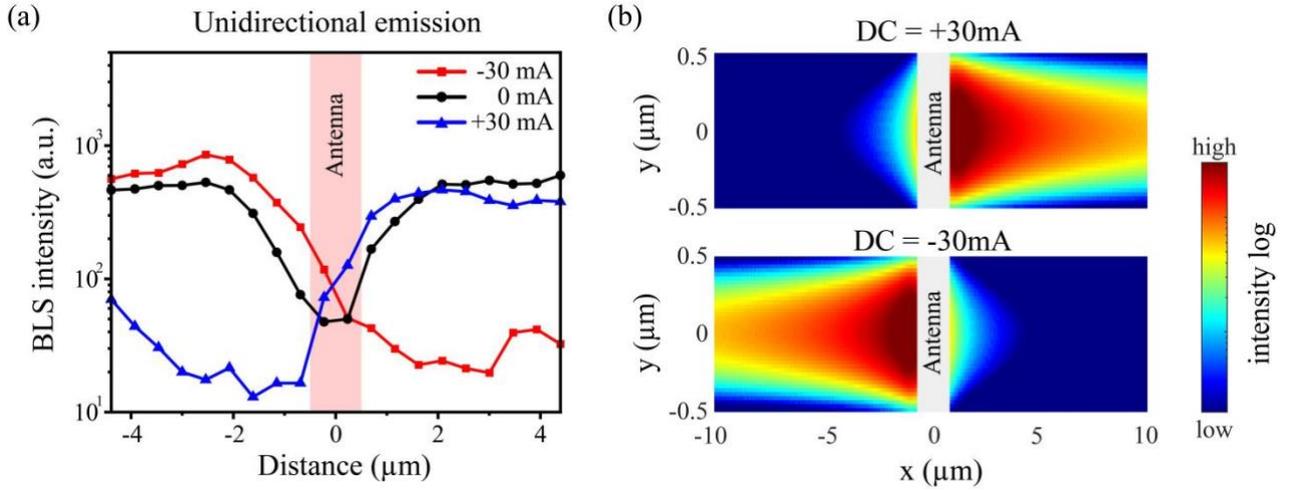

*Figure 3. **Spatial distribution and simulated 2D spin-wave intensity map.** (a) The spin-wave intensity is measured using μBLS in a rectangular region symmetric about the antenna on the waveguide, with averaging performed along the width direction. An RF signal of frequency 4.8 GHz (-5 dBm) with different DC signals of ±30mA, 0mA are both applied to the antenna. (b) Simulation results of unidirectional magnon emission, where the color represents the spatial distribution of spin-wave intensity (logarithmic scale). The direction of unidirectional emission can be changed by switching the direction of the DC signals.*

To demonstrate unidirectional spin-wave emission in real space, a microwave signal at a frequency of 4.8 GHz in the nonreciprocity region (see Fig. 2c) with an RF power of -5 dBm is applied to the antenna. Microfocused BLS is used to scan an area of $18.5 \times 1.1$ μm² with $20 \times 5$ points symmetric to the antenna. Figure 3(a) shows the spin-wave intensity averaged over the width as a function of length for different cases. The pink region indicates the position of the antenna with a width of 1 μm. The BLS intensity drops in the center because the laser signal is blocked by the non-transparent gold-made antenna. In the absence of a DC signal, the propagation behavior of the spin waves on the waveguide is symmetric around the antenna, as shown by the black curve. When DC signals are applied, the spin waves have a preferred direction of propagation and this direction can be switched by changing the direction of the DC signal. This unidirectional excitation is also confirmed by micromagnetic simulations. Figure 3(b) shows the simulated 2D spin-wave intensity of the excitation frequency 4.64 GHz for different conditions (see details of the simulation in Methods and Supplementary Materials).

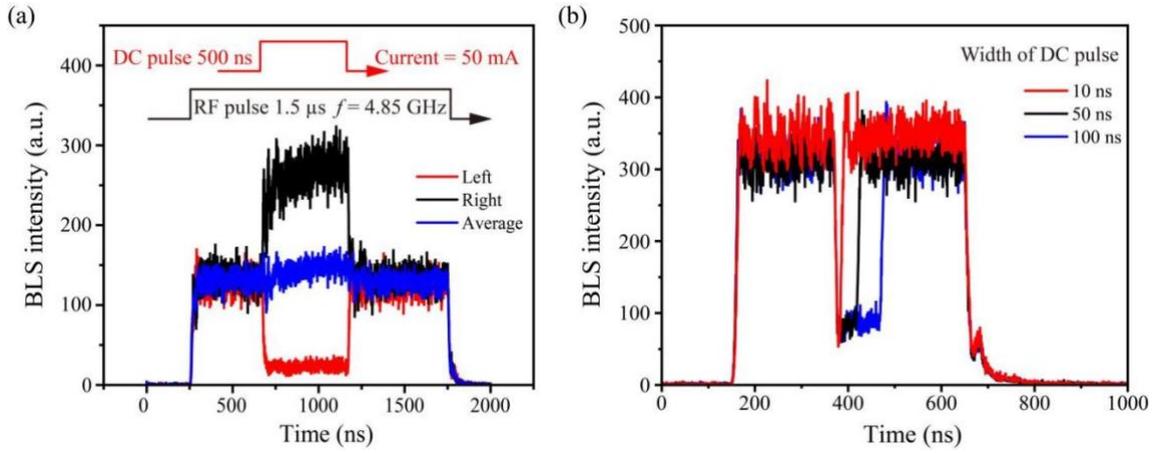

*Figure 4. **Time-resolved BLS measurements**. (a) BLS intensity as a function of time. Within a 2 μs cycle, a 1.5 μs microwave pulse with a frequency of 4.85 GHz is applied. A DC pulse of 0.5 μs width is added in the middle of the cycle. (b) The similar experiments in (a) with reduced DC pulse widths of 100 ns, 50 ns, and 10 ns. The results demonstrate that nanosecond switching can be achieved experimentally.*

**Nanosecond switchable unidirectional magnon emitter.** As mentioned above, this nonreciprocity is caused by the DC-induced asymmetric Oersted field, which provides a way to realize the fast-switching feature. To experimentally observe the fast switching, a constant DC is replaced by a pulsed DC and added in the middle of the RF pulse, as shown in the inset of Fig. 4(a). In each 2 μs cycle, a 1.5 μs microwave pulse of 4.85 GHz frequency is applied, followed by the insertion of a 0.5 μs DC pulse in the middle of the cycle. The time-resolved BLS signal is measured at symmetric positions on both sides of the waveguide as shown by the red and black curves. It is clear that the spin wave on the left side is suppressed due to the potential barrier created by the DC pulse. In contrast, the spin wave on the right side is enhanced by the reflection waves from the left side. The averaged intensity (blue curve) is nearly constant indicating energy conservation. Another advantage of this method is that the DC pulse does not change the total energy pumped from the microwave to the spin wave. It only redistributes the energy for different propagation directions by creating a potential barrier/well and finally increasing the excitation efficiency. This behavior is crucial for the potential application of magnonic devices with controlled modulation of spin-wave signals.

In addition, to show how fast the switch can be, we greatly reduce the width of the DC pulse to 10 ns. Experimental measurements are performed using µBLS on the suppressed side of the waveguide and the results are shown in Fig. 4(b). The spin-wave intensity is significantly suppressed during the DC pulses even for the pulse width of 10 ns, demonstrating that our unidirectional excitation method achieves a switchable magnon emitter on a nanosecond timescale. This fast switching is also repeated in the micromagnetic simulations [see Supplementary Materials]. Unlike conventional approaches, which either lack direction switching capability or require longer switching times (e.g., reversal of the magnetization direction in microseconds), our method allows rapid direction switching within nanoseconds. This highlights the significant advantage of our unidirectional excitation technique in enabling fast and efficient control of spin waves.

**Discussion**

In this study, we present a novel approach to realizing a fast, switchable unidirectional magnon emitter using an asymmetric Oersted field generated by a DC signal in forward volume spin waves geometry. This technique addresses the limitations of traditional Damon-Eshbach (DE) mode-based methods, which are typically directionally fixed and require significant time to switch. The key innovation of our method is the ability to induce in a single integrated device an asymmetric DC field that creates spatially asymmetric potential landscape - a well and a barrier for opposite propagation directions, allowing for effective unidirectional excitation. This asymmetry enables for been rapidly switched by simply reversing the DC, an improvement over conventional methods that often require slower magnetic field adjustments. Experimental results demonstrate that our technique achieves rapidly switchable magnon emitters on the nanosecond timescale, with the spin-wave signal intensity varying significantly between different sides of the waveguide. We believe that this rapid switching capability is critical for the development of high-speed magnonic networks, providing precise control over signal propagation and increasing the overall efficiency of the system.

**Methods**

**YIG growth by liquid phase epitaxy.** The YIG thin film is grown on top of a 500 µm-thick (111) gadolinium gallium garnet (GGG) substrate by liquid phase epitaxy (LPE) [49]. The parameters of the unstructured thin film have been characterised by stripline vector-network-analyser ferromagnetic resonance spectroscopy and BLS spectroscopy and yield a saturation magnetisation of $M_s = (140.7 \pm 2.8)$ kA/m, Gilbert damping parameter $\alpha = (1.75 \pm 0.08) \times 10^{-4}$, inhomogeneous linewidth broadening $\mu_0 \Delta H_0 = (0.18 \pm 0.01)$ mT, and exchange constant $A_{ex} = (4.22 \pm 0.21)$ pJ/m. These parameters are typical for high-quality thin YIG films [48,49].

**Nanoscale waveguide fabrication.** The YIG waveguides are fabricated by chromium mask and argon ion beam etching. On a clean YIG sample, the adhesive layer, positive resist (CSAR) and conductive layer (ELECTRA) were spin-coated and baked. The waveguides were then patterned using an e-beam lithographer (30 keV, 100 pA). After patterning, the sample was placed in water to remove the conductive layer and in CSAR developer. The resist residues from the developed areas were removed by oxygen etching. On the developed sample, 250 nm of chromium was deposited by electron beam physical vapor deposition. After lift-off procedure, the YIG waveguides were etched using argon ion beam (300W, 0.16mA/cm$^2$). After etching, the chromium mask still on the YIG waveguides were dissolved with Cr solvent. In the following step, we fabricated microwave antennas on the YIG waveguides using e-beam lithography and electron-beam physical vapour deposition (10 nm Ti, 320 nm Cu, 20 nm Au).

**BLS measurements.** A single-frequency laser with a wavelength of 457 nm is used, focused on the sample using a microscope objective (magnification 100× and numerical aperture N.A.=0.8). The laser power of 2.8 mW is focused on the sample. A uniform out-of-plane external field of 320 mT is provided by a NdFeB permanent magnet with a diameter of 70 mm. Microwave signals with different powers and frequencies were applied to the antenna to excite propagation spin-waves.

**Micromagnetic simulations.** The micromagnetic simulations were performed by the GPU-accelerated simulation package Mumax$^3$, including both exchange and dipolar interactions, to calculate the space- and time-dependent magnetisation dynamics in the investigated structures [50]. The parameters of a nanometre-thick YIG film were used [48,49]: saturation magnetisation $M_s = 1.407 \times 10^5$ A/m, exchange constant $A = 4.2$ pJ/m. The Gilbert damping is increased to $\alpha = 2 \times 10^{-4}$ to account for the inhomogeneous linewidth which cannot be directly plugged into Mumax$^3$ simulations. The Gilbert damping at the end of the device was set to exponentially increase to 0.5 to avoid spin-wave reflection. The mesh was set to 20×25×97 nm$^3$ (single layer along the thickness) for YIG waveguide. An external field $B_{ext} = 320$ mT is applied along the out-of-plane axis (z-axis as shown in Fig. 1) and thus sufficient to saturate the structure in this direction.

To excite propagating spin-waves, we first calculate the Oersted field distribution of a 1 μm wide strip antenna with current of 10 mA in the magneto-static approximation and plug it into Mumax$^3$ with a varying microwave frequency $f$. The $M_y(x,y,t)$ of each cell was collected over a period of 100 ns and recorded in 50 ps intervals.

**Acknowledgements**

**Funding:** The project is funded by the National Key Research and Development Program of China (Grant No. 2023YFA1406600), the Austrian Science Fund (FWF) via Grant No. I 4917-N (MagFunc) and Grant No. F65 (SFB PDE), the European Research Council (ERC) Starting Grant 101042439 "CoSpiN" and the Deutsche Forschungsgemeinschaft (DFG, German Research Foundation) – 271741898 and TRR 173 - 268565370 ("Spin + X", Project B01). Q. W. was supported by the startup grant of Huazhong University of Science and Technology Grants No. 3034012104. R.V. acknowledges support by the Ministry of Education and Science of Ukraine, project # 0124U000270. We acknowledge CzechNanoLab Research Infrastructure supported by MEYS CR (LM2023051).

**Author contributions:** M. G., Y. W., X. G. performed BLS measurements. Y. W. carried out micromagnetic simulations. K. D. fabricated the nanoscale YIG waveguides. R. V. developed the analytical theory. C. D. grew the YIG film. Q. W. conceived the idea and led this project. Y. W., Q. W. wrote the manuscript with the help of all the coauthors. All authors contributed to the scientific discussion and commented on the manuscript.

**Competing interests:** The authors declare no competing interests.

**Data and materials availability:** All data needed to evaluate the conclusions in the paper are present in the paper and/or the Supplementary Materials.

**Correspondence** and requests for materials should be addressed to Q. W.